\documentclass[traditabstract]{aa} 

\usepackage{txfonts}
\usepackage{graphicx}
\usepackage{natbib}
\usepackage{color}

\newcommand{\oex}{\mbox{\object{EX\,Hya}}}

\newcommand{\lyalp}{Ly$\alpha$}

\newcommand{\porb}{$P_\mathrm{orb}$}
\newcommand{\pspin}{$P_\mathrm{spin}$}
\newcommand{\gs}{g\,s$^{-1}$}

\newcommand{\kms}{km\,s$^{-1}$}

\newcommand{\nh}{$N_\mathrm{H}$}

\newcommand{\nhint}{$N_\mathrm{H,int}$}
\newcommand{\nhist}{$N_\mathrm{H,ist}$}
\newcommand{\atoms}{H-atoms\,cm$^{-2}$}
\newcommand{\fpc}{$f_\mathrm{pc}$}
\newcommand{\teff}{$T_\mathrm{eff}$}

\newcommand{\ktsh}{k$T_\mathrm{sh}$}

\newcommand{\erg}{erg\,s$^{-1}$}
\newcommand{\ergcm}{erg\,cm$^{-2}$}
\newcommand{\ergs}{erg\,cm$^{-2}$s$^{-1}$}

\newcommand{\ergskev}{erg\,cm$^{-2}$s$^{-1}$keV$^{-1}$}

\newcommand{\ten}[2]{#1\,\times\,10^{#2}}

\newcommand{\hsh}{$h_\mathrm{sh}$}
\newcommand{\rsun}{$R_\odot$}
\newcommand{\msun}{$M_\odot$}
\newcommand{\msunyr}{$M_{\odot}$yr$^{-1}$}

\usepackage{amstext}

\begin{document}

\title{Current and secular accretion rates of EX Hydrae}

\author{K.~Beuermann\inst{1} \and K.~Reinsch\inst{1}}
 
\institute{ 
  Institut f\"ur Astrophysik und Geophysik, Georg-August-Universit\"at,
  Friedrich-Hund-Platz 1, D-37077 G\"ottingen, Germany }

\date{Received  24 April 2024; accepted 9 May 2024}

\abstract{We report an observed accretion rate of\ 
  $\dot M_1\!=\!\ten{(3.86\pm0.60)}{-11}$\,\msunyr\   for the white dwarf in
  the short-period, intermediate polar \oex. This  result is based upon the
  accretion-induced $4\pi$-averaged energy flux from $2.45\mu$m to
  100\,keV and the corresponding luminosity at the Gaia distance of
  56.77\,pc. Our result is in perfect agreement with the theoretical mass
  transfer rate from the secondary star induced by gravitational
  radiation (GR) and the spin-up of the white dwarf,
  $-\dot M_2\!=\!\ten{(3.90\pm0.35)}{-11}$\,\msunyr;  24\% of it is 
  caused by the spin-up. The agreement indicates that mass transfer is
  conservative. The measured $\dot M_1$ obviates the need for angular
  momentum loss (AML) by any process other than GR. We complemented this
  result with an estimate of the mean secular mass transfer rate over
  $\sim\!10^7$\,yr by interpreting the non-equilibrium radius of the
  secondary star in \oex\ based on published evolutionary
  calculations. This suggests a time-averaged mass
  transfer rate enhanced over GR by a factor
  $f_{\mathrm{GR}}\!\gtrsim\!2$.  Combined with the present-day lack
  of such an excess, we suggest that an enhanced secular AML is due
  to an intermittently active process, such as the proposed frictional
  motion of the binary in the remnants of nova outbursts. We argue
  that \oex, despite its weakly magnetic nature, has evolved in a very
  similar way to non-magnetic CVs. We speculate that the discontinuous
  nature of an enhanced secular AML may  similarly apply to the latter. }

   
\titlerunning{Accretion rates of \oex}
\authorrunning{K.~Beuermann \& K.~Reinsch}

\maketitle


\section{Introduction}
\label{sec:intro}

Mass transfer in cataclysmic variables (CV) is driven by angular
momentum loss (AML) from the binary orbit. The processes considered in
the classical models of CV evolution are magnetic braking at long
orbital periods, \porb\,$>\!3$\,h, and gravitational radiation (GR)
below the $2\!-\!3$\,h period gap \citep[e.g.,][]{rappaportetal82}.
However, it has become clear that numerous parameters that
characterize the CV population are not satisfactorily described by the
standard model of binary population studies \citep[e.g.,][henceforth
KBP11]{kniggeetal11}, among them the degree of bloating of the
secondary stars as they are driven out of thermal equilibrium and the
effective temperatures of the compressionally heated accreting white
dwarfs (WDs). Some of these deficiencies could be remediated by
adjusting the relative strength of the secular AML processes below and
above the period gap
\citep[KBP11,][]{schreiberetal16,palaetal17,palaetal22,bellonietal18,mcallisteretal19}.
Specifically, below the gap, KBP11 raised the systemic AML from
$\dot J_\mathrm{GR}$ by GR alone to an average
$\dot J_\mathrm{sys}\!=\!2.47\times\dot J_\mathrm{GR}$, which
characterizes the revised (optimal) model of KBP11. The enhanced mass
transfer not only fits the bloated radii of the secondary stars, but
it also solved several previously unexplained issues related to binary
evolution, greatly increasing the credibility of the approach. The
long-term effects are similar for an AML that occurs continuously or
in repetitive bursts, as long as the duty cycle of the bursts is
shorter than the timescale of the process considered; for
instance, the radius inflation.  The culprit identified by
\citet{schreiberetal16} is frictional motion of the binary in the
shells of slow nova outbursts.  These events take place in short
period CVs roughly every $10^6$\,years.  This is short enough for all
secondaries to share the same radial evolution, but possibly too long
a timescale to completely homogenize the temperatures of the
compressionally heated WDs.

In this work, we endeavor to shed some light on the origin of the missing AML for
the case of the intermediate polar (IP) EX Hya. This disk-accreting
binary with an orbital period of 98\,min equals in many respects a
standard non-magnetic CV, except that its primary is weakly magnetic
with a surface field strength $B\!\simeq\!0.35$\,MG
\citep[][henceforth BR24]{beuermannreinsch24}. We anticipate that it
evolves similarly to a non-magnetic CV. Our approach involves an accurate
measurement of the current accretion rate of \oex\ via its
luminosity that is valid over the last few decades. We confront it
with an estimate of the larger secular mean accretion rate obtained
from the inflated radius of its secondary star, valid for the last
$\sim\!10^7$\,yr.

\section{Conservative mass transfer with WD spin-up}
\label{sec:cons}

In this work, we employ the description of conservative mass transfer by
\citet{ritter85}, which  includes the variation of the spin angular
momenta of WD and secondary star (his Eqs.~1 and A19). The mass
transfer rate $-\dot M_2\!=\!\dot M_1$ is obtained by equating the
systemic angular momentum loss of the binary to the sum of the
$\dot M_1$-dependent temporal changes of the orbital and spin angular
momenta of the stellar components,
$\dot J_\mathrm{sys}\!=\dot J_\mathrm{orb}+\dot J_\mathrm{S1}+\dot
J_\mathrm{S2}$ (index 1 for primary and 2 for secondary). Solving for
$\dot M_1$ , we obtain:
\begin{equation}
  \dot M_1=\frac{S_1-\dot J_\mathrm{sys}}{\mathrm{G}^{2/3}[M_1(5+3\alpha_2)/6-M_2](M\omega)^{-1/3}+S_3-S_2}~,
\label{eq:ritter1}
\end{equation}
with $M\!=\!M_1\!+\!M_2$ and the spin-related terms as:
\begin{eqnarray}
S_1 & = & (r_\mathrm{g1}R_1)^2M_1\dot \Omega_1,\\
S_2 & = & (r_\mathrm{g1}R_1)^2\Omega_1(1+2\alpha_1),\\
S_3 & = & (r_\mathrm{g2}R_2)^2\omega\,(3+\alpha_2)/2.
\label{eq:ritter2}
\end{eqnarray}
We equate $\dot J_\mathrm{sys}$ to $f_\mathrm{GR} \dot J_\mathrm{GR}$, where
\begin{equation}
\dot J_\mathrm{GR}=\frac{32\mathrm{G}^{7/3}}{5c^5}M_1^2M_2^2M^{-2/3}\omega^{7/3},
\label{eq:ritter3}
\end{equation}
is the AML by GR and $f_\mathrm{GR}\ge1$ accounts for any additional
currently active AML process. Of the spin-related terms, $S_1$
describes the accretion of angular momentum by the WD, while $S_2$ and
$S_3$ are the effects of the mass transfer-related internal changes of
the WD and the companion star, respectively. The quantities $P_1$,
$\Omega_1\!=\!2\pi/P_1$, and
$\tau_\mathrm{spin}\!=\!P_1/\dot P_1\!=\!\Omega_1/\dot \Omega_1$ are
the spin period, angular frequency, and spin-up timescale of the WD,
respectively, while $\omega=\Omega_2$ is the orbital and spin angular
frequency of the synchronously rotating secondary star. Then,
$r_\mathrm{g1}$ and $r_\mathrm{g2}$ are the radii of gyration and
$\alpha_1$ and $\alpha_2$ the exponents of the power-law mass-radius
relations, $R_1\propto M_1^{\,\alpha_1}$ and
$R_2\propto M_2^{\,\alpha_2}$, for the WD and secondary star,
respectively.  The mass transfer is significantly enhanced by the loss
of orbital angular momentum to spin angular momentum of the WD, but is
driven solely by the systemic AML by GR or another process
\citep{ritter85}. The remaining uncertainty in the calculated
accretion rate is determined by the observational errors of
$M_2, \alpha_2$, $M_1$, and $\tau_\mathrm{spin}$, in this order.

We go on to derive the binary ephemeris expected for a constant
spin-up rate $\dot \Omega\!=\!q$ of the WD (Eq.~2) and the related
spin-up timescale, $\tau_\mathrm{spin}$, of the WD. For simplicity, we
drop here the subindex 1, indicating the WD. We start from
d$\Omega\!=\!q$d$t$ and d$t\!=\!P$d$E$, with $E$ a continuous version
of the spin cycle number, obtaining $\Omega(E)$ by integration, $P(E)$
by expansion of $2\pi/\Omega(E)$, and the ephemeris by integration of
$P(E)$. This leads to $\Omega^2\!=\!\Omega_0^2+4\pi qE$ and to
$P\!=\!P_0+2CE+3DE^2$ plus higher terms, with $\Omega_0$ and $P_0$ the
values at $E\!=\!0$, $C\!=-qP_0^3/(4\pi)$, and
$D\!=\!2q^2P_0^5/(4\pi)^2$. Integration of d$t\!=\!P$d$E$ starting
with $t=T_0$ at $E\!=\!0$ gives the ephemeris $T=T_0+P_0E+CE^2+DE^3$
plus higher terms. Since $D$ is a multiple of $C$, and all terms from
the cubic on are minute, the ephemeris is effectively quadratic. The
spin-up timescale is
$\tau_\mathrm{spin}(E)\!=\!-P/\dot
P\!\simeq\!-P_0^2/(2C)+2P_0E\simeq\!-P_0^2/(2C)$, where $2P_0E$
describes the small variation of $\tau_\mathrm{spin}$ with $E$.

\begin{table}[b]
  \caption{Spin-maximum times measured from AAVSO light curves, using
    the tool LCGv2. The data were taken between 2003 and 2023, mostly
    in the Johnson $V$-band. The timing error is estimated at
    typically 5 min. The complete table is available at CDS
    Strasbourg. }
\begin{tabular}{@{\hspace{1mm}}c@{\hspace{4.0mm}}c@{\hspace{4.0mm}}c@{\hspace{4.0mm}}c}
\hline \hline \noalign{\smallskip}
BJD(TDB)    & BJD(TDB)    & BJD(TDB) & BJD(TDB) \\ 
\noalign{\smallskip} \hline
  \noalign{\medskip}
52739.27482  & 52739.36905  &  52743.27500  &  52743.36678 \\ 
52746.25106  & 52746.34548  &  52746.44288  &  52746.53441 \\
\ldots       & \ldots       &  \ldots       &  \ldots      \\
\ldots       & \ldots       &  \ldots       &  \ldots      \\
60092.61082  & 60093.58639  &  60093.63143  &  60093.72461 \\ 
60097.68124  & 60106.66837  &  60108.57066  &  60108.67018 \\[1.0ex]
\noalign{\medskip} \hline
\end{tabular}
\label{tab:spinmax}
\end{table}

\subsection{Spin-up timescale of the WD}
\label{sec:spinup}

\begin{figure}[t]
\includegraphics[height=89mm,angle=270,clip]{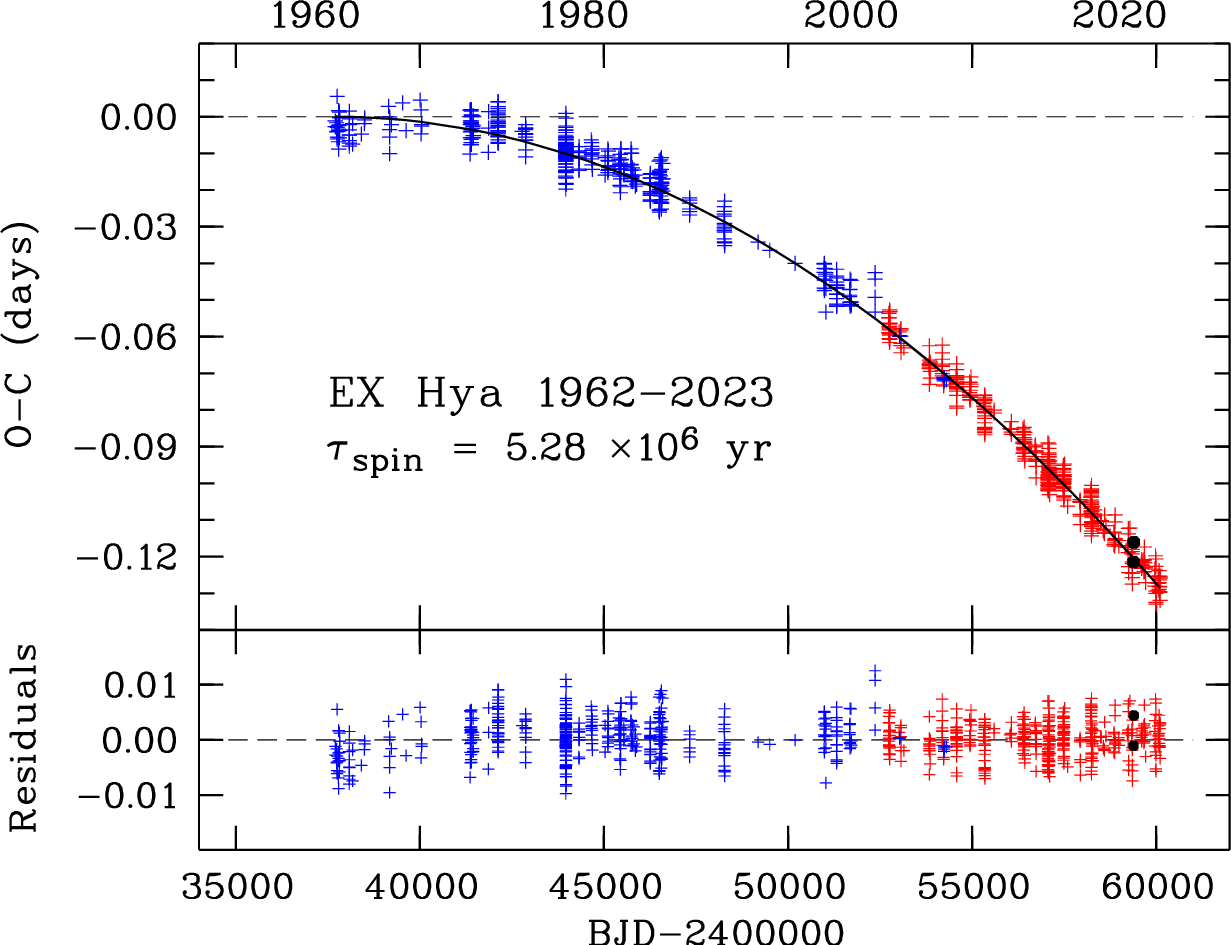}
\caption{Quadratic term of the parabolic ephemeris of the spin maximum
  times of \oex\ (top). Blue: data from \citet{maucheetal09}. Red and
  black dots: this work (Table~1, see text).  Residuals from the
  parabolic fit (bottom).}
\label{fig:tau}
\end{figure}
 

\citet{kruszewskietal81} discovered the secular decrease of the spin
period of \oex. The initially suggested values of $\tau_\mathrm{spin}$
around 3\,Myr were corrected upward to around 5\,Myr as more data were
accumulated \citep{maucheetal09,andronovbreus13}. We measured 403
additional spin-maximum times from AAVSO light curves, mostly in the
Johnson $V$-band, using the online tool
LCGv2\,\footnote{https://www.aavso.org/LCGv2/}
(Table~\ref{tab:spinmax}).  Of these, a small fraction overlaps with
the timings of \citet{breusandronov13}.  Two independent spin-maximum
times were obtained from light curves taken in 2021 at the Merrimack
College \citep{duston23}.  We accepted the errors of the mixed bag of
435 spin maximum times collected by \citep{maucheetal09}, except for
ten entries with reported errors down to 4.3~s, which we raised to
1~min or 0.00070 d. The uncertainties of all 403 new timings were
estimated at 5.0~min or 0.00347~d. We fit the combined data with the
quadratic ephemeris, $T_\mathrm{max}\!=\!T_0+P_0E+CE^2$, and obtained
the spin-up timescale from $P_0$ and $C$.  The fit gives
\begin{eqnarray}
  T_\mathrm{max}\!&\!=\!&\!2437699.89333\!+\!0.0465464497 E\!-\!5.61\!\times\!10^{-13}\,E^2 ,\\[-1.0ex]
\label{eq:tau}
                  &   &\hspace{18.0mm}11\hspace{21.0mm}4\hspace{9.5mm}1  \nonumber
\end{eqnarray} 
with $\chi^2\,=\,824.4$ for 835 degrees of freedom (dof).  The upper
panel of Fig.~\ref{fig:tau} shows the $O\!-\!C$ values in
BJD(TDB)\,\footnote{\citet{eastmanetal10}:~https://astroutils.astronomy.osu.edu/}
relative to the linear part of the ephemeris, with Mauche's data in
blue, the AAVSO data in red, and the Merrimack points as black
dots. The lower panel shows the residuals from the parabolic fit, with
rms values of 5.3~min and 4.4~min for Mauche's data and the AAVSO
data, respectively. The spin-up timescale at $E=0$ is
$\tau_\mathrm{spin}\!=\!-P_0^2/(2C)\!=\!\ten{5.28}{6}$\,yr, resulting
in a decrease of the spin period by 47\,ms between 1962 and 2023. The
quadratic fit is not perfect, displaying small systematic offsets
between model and data\,: it is high by a couple of minutes for the
subset of the first 37 maxima observed in the mid-1960s and low by a
bit less than a minute for a larger group of data around 1980. This
led \citet{maucheetal09} to advocate for a cubic fit. When employed to
the present data set, it gives an improved $\chi^2\,=\,787.2$ for 834
dof; this solution appears to justify the additional parameter. Their
fit describes the present data well, but implies a diverging
$\tau_\mathrm{spin}$ within 100\,yr into the future.
We consider it more likely that variations in the skewness of the spin
light curves and/or differences in the methods of localizing the
maxima in noisy light curves are responsible for the $O-C$ offsets of
subgroups of the data. We estimate that the true error of parameter
$C$ is rather $\ten{0.10}{-13}$, which gives
$\tau_\mathrm{spin}\!=\!\ten{(5.28\!\pm\!0.10)}{6}$\,yr.

\subsection{System parameters}
\label{sec:system}


\begin{table}[b]
\caption{System parameters of EX Hya }
\begin{tabular}{@{\hspace{0.0mm}}l@{\hspace{2.0mm}}l} 
  \hline \hline \noalign{\smallskip}
  Parameter &  Value \\[0.5ex]
  \noalign{\smallskip} \hline
  \noalign{\medskip}
  Distance $d$ (pc)                       &  $56.77\pm 0.05^{~a)}$               \\
  Orbital period \porb\ (s)                      &  5895.4                      \\ 
  Spin period \pspin\ (s)                         &  4021.6                      \\ 
  Velocity amplitude of WD $K_1$ (\kms)   &  $58.9\pm1.8^{~b)}$           \\
  Velocity amplitude of secondary $K_2$ (\kms) &  $432.4\pm 4.8$$^{~c)}$              \\
  Mass ratio $q=M_2/M_1$                    &  0.1362$\pm 0.0044$       \\
  Duration of X-ray eclipse $t_{\mathrm{ecl}}$  (s)  & $157\pm4$$^{~d)}$  \\
  Inclination $i$ (\degr)                     &  $78.0\pm0.2$      \\ 
  Minimal inclination $i_0$ for WD center (\degr)&  $78.1\pm0.1$      \\ 
  Total mass $M\!=\!M_1\!+\!M_2$ (\msun)                  &  $0.895\pm 0.028$            \\ 
  Mass of WD $M_1$  (\msun)               &  $0.788\pm 0.025$            \\
  Radius of WD $R_1$ ($10^8$\,cm)         &  $7.35\!\pm\!0.23\,^{~c)}$           \\
  Mass of secondary $M_2$ (\msun)         &  $0.1074\pm 0.0047$            \\
  Radius of secondary $R_2$ (\rsun)       &  $0.1513\pm 0.0022$          \\
  Separation of stars $a$ ($10^{10}$ \,cm) &  $4.712\pm0.050$         \\
  Spin-up timescale $\tau_\mathrm{spin}$ of WD ($10^6$\,yr) & $5.28\!\pm\!0.10$\\[0.5ex]
  \noalign{\smallskip} \hline                                    
\end{tabular}

\smallskip $^{a)}$ Gaia distance \citep{bailerjonesetal21}. $^{b)}$
Weighted mean from BR24, see text. $^{c)}$ From BR08. $^{d)}$ From \citet{mukaietal98}.\\
\label{tab:system}
\end{table}

The component masses were determined by the velocity amplitudes $K_1$
and $K_2$ of the stellar components and the inclination, $i$, obtained
from the duration $t_\mathrm{ecl}$ of the grazing X-ray eclipse of the
lower pole of the WD \citep{beuermannosborne88}.  We assumed that the
centroid of the X-ray emission is located close to the WD surface,
noting that a substantial elevation could lower the derived
inclination by up to $0.8$\degr.  Table~\ref{tab:system} provides an
update of the system parameters presented in BR24 and
\citet[][henceforth BR08]{beuermannreinsch08}. The value of $K_1$ is
a weighted mean of the measurements of \citet{belleetal03},
\citet{hoogerwerfetal04}, \citet{echevarriaetal16}, and BR24,
$K_1\!=\!58.9\!\pm\!1.8$\,\kms, while $K_2\!=\!432.4\pm4.8$\,\kms\
is from BR08. The well-defined FWHM of the partial, flat-bottomed
$3-15$\,keV X-ray eclipse is $t_\mathrm{ecl}=157\pm4$\,s
\citep{mukaietal98}. We used the three-dimensional (3D) description of
the Roche geometry of \citet{kopal59}\,\footnote{The description of
  \citet{sirotkinkim09}, which is based on $n\!=\!1.5$ polytropes
  yields only minimally different results.}. The quoted values of
$K_1, K_2$, and $t_\mathrm{ecl}$ translate into the parameters listed
in Table~\ref{tab:system}. The inclination, $i,$ and the limiting
inclination $i_0$ for no eclipse refer to the center of the WD. Both
agree closely indicating that, in fact, the shadow of the secondary
star cuts right through the WD.

Our measured primary mass of 0.788\,\msun\ is sufficiently close to
the 0.75\,\msun\ adopted by KBP11 in their evolutionary model
calculations to allow their application to \oex. The standard and
revised model tracks, with AML by GR only and a best-fit enhanced mean
secular AML with $f_\mathrm{GR}\!=2.47\pm0.22$, respectively, are
tabulated as functions of $M_2$ in their Tables $3-8$. The revised
model provides a best fit to the data, while the standard model fails
by 6.6~$\sigma$ (standard deviations) in $f_\mathrm{GR}$.  For the
secondary star in \oex, with a mass of 0.1074\,\msun, the standard and
revised models predict radii $R_2\!=\!0.1451$\,\rsun\ and
0.1512\,\rsun, respectively, at the same mass. Its observed radius
of $0.1513\!\pm\!0.0022$\,\rsun\ coincides with that of the revised
model and formally exceeds that of the standard model by
$2.8\,\sigma$. The agreement in $R_2$ suggests that this mildly
magnetic, disk-accreting IP evolved similarly to the sample of
non-magnetic CVs and experienced a similarly enhanced secular mass
transfer rate. The rates predicted by the standard and revised model
tracks are $-\dot M_2\!=\!\ten{3.13}{-11}$\,\msunyr\ and
$\ten{6.69}{-11}$\,\msunyr, respectively (increasing by a factor of
2.14). We expect the latter to represent the mean secular mass
transfer rate in \oex\ as well.  Given the small difference in $R_2$,
  it is necessary to consider the systematic uncertainties of the
  approach, to which we return in Sect.~\ref{sec:disc}.
  
\begin{figure}[t]
\hfill
  \includegraphics[height=89mm,angle=270,clip]{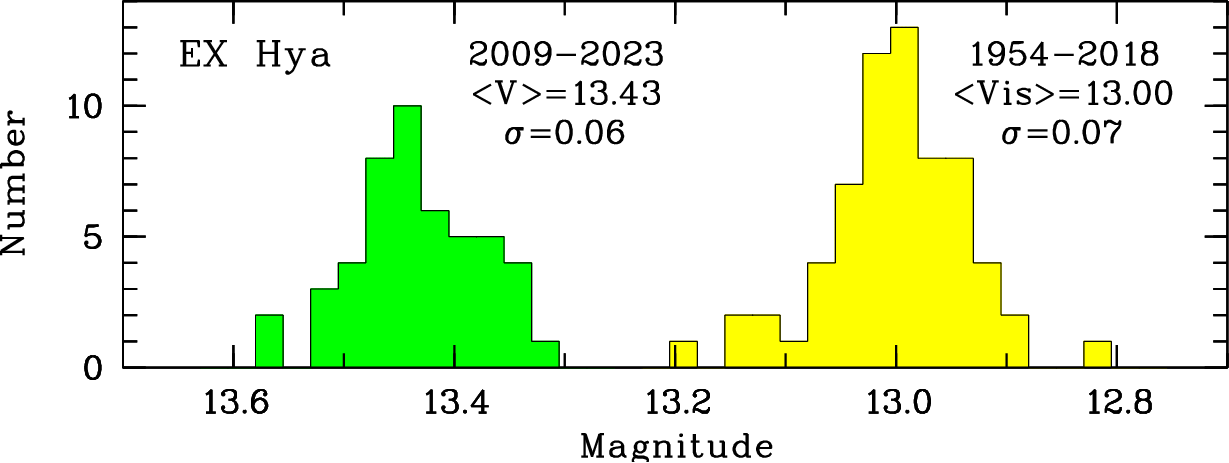}
  \caption{Frequency distribution of the yearly mean AAVSO visual
    magnitudes of \oex\ outside outbursts (right) and the about monthly mean
    Johnson~V magnitudes (left), displaying the low degree of
    variability.}
  \label{fig:vis}
\end{figure}

\subsection{Theoretical mass transfer rate $-\dot M_\mathrm{2}$ } 
\label{sec:trans}

We calculated the theoretical mass transfer rate \mbox{from Eqs.~1
  to~5}, using the system parameters of Table~\ref{tab:system},
$\alpha_1\!\simeq\!-1.0$ for the heated WD, and the radii of gyration
$r_\mathrm{g1}\!=\!0.423$ and $r_\mathrm{g2}\!=\!0.453$ for a
polytrope of index 1.5 from \citet{ritter85}. We adopted the values of
$\alpha_2$ from the evolutionary tracks of KBP11 at a mass of
0.1074\,\rsun: $\alpha_2\!= \!0.764$ for the standard model and
$\alpha_2\!= \!0.690$ for the enhanced AML of the revised model (their
Tables 5 and 6). For $f_\mathrm{GR}\!=\!1.0$, equivalent to the
standard model, and $\dot \Omega_1\!=\!0$, the transfer rate is
$-\dot M_\mathrm{2,GR}\!=\!\ten{(2.98\!\pm\!0.27)}{-11}$\,\msunyr, in
reasonable agreement with the above $\ten{3.13}{-11}$\,\msunyr\ from
KBP11. With the observed
$\tau_\mathrm{spin}\!=\!\ten{(5.28\!\pm\!0.10)}{6}$\,yr, the transfer
rate calculated from Eq.~\ref{eq:ritter1} rises to
$\ten{(3.90\!\pm\!0.36)}{-11}$\,\msunyr, a sizable increase by
31\%. The quoted error of the GR-driven rate arises from the
observational uncertainties in $M_2, \alpha_2,$ and $M_1$, in this
order. The smaller relative error of the spin-up contribution comes
from $R_1, M_1$ and least from $\tau_\mathrm{spin}$.  With
$\alpha_2\!=\!0.690$ for the revised model track of KBP11, the target
rate of $-\dot M_2\!=\!\ten{6.69}{-11}$\,\msunyr\ is obtained for
$f_\mathrm{GR}\!=\!2.21$ and $f_\mathrm{GR}\!=\!1.88$, without and
with the spin-up term, respectively.

\section{Observational accretion rate  $\dot M_\mathrm{1}$ }
\label{sec:obs}

In this section, we detail the tedious procedure of deriving the
accretion rate of the WD from the overall spin-averaged spectral
energy distribution (SED) of \oex. Apart from the secondary star and
the WD in its inactive state, all emission components are
accretion-induced. Piecing together the SED from non-simultaneous
observations is feasible because of the observed lack of long-term
variability, which is well documented in the optical region and
indicated also for the X-ray regime. The long-term mean $V$-band flux
is used as an anchor-point for the spin-averaged SED and the
corresponding integrated energy flux. Obtaining the accretion
luminosity, $L_\mathrm{acc}$, requires the conversion of the
spin-averaged to the $4\pi$-averaged energy flux
(Sect.~\ref{sec:4pi}).

\begin{figure*}[t]
\includegraphics[height=106mm,angle=270,clip]{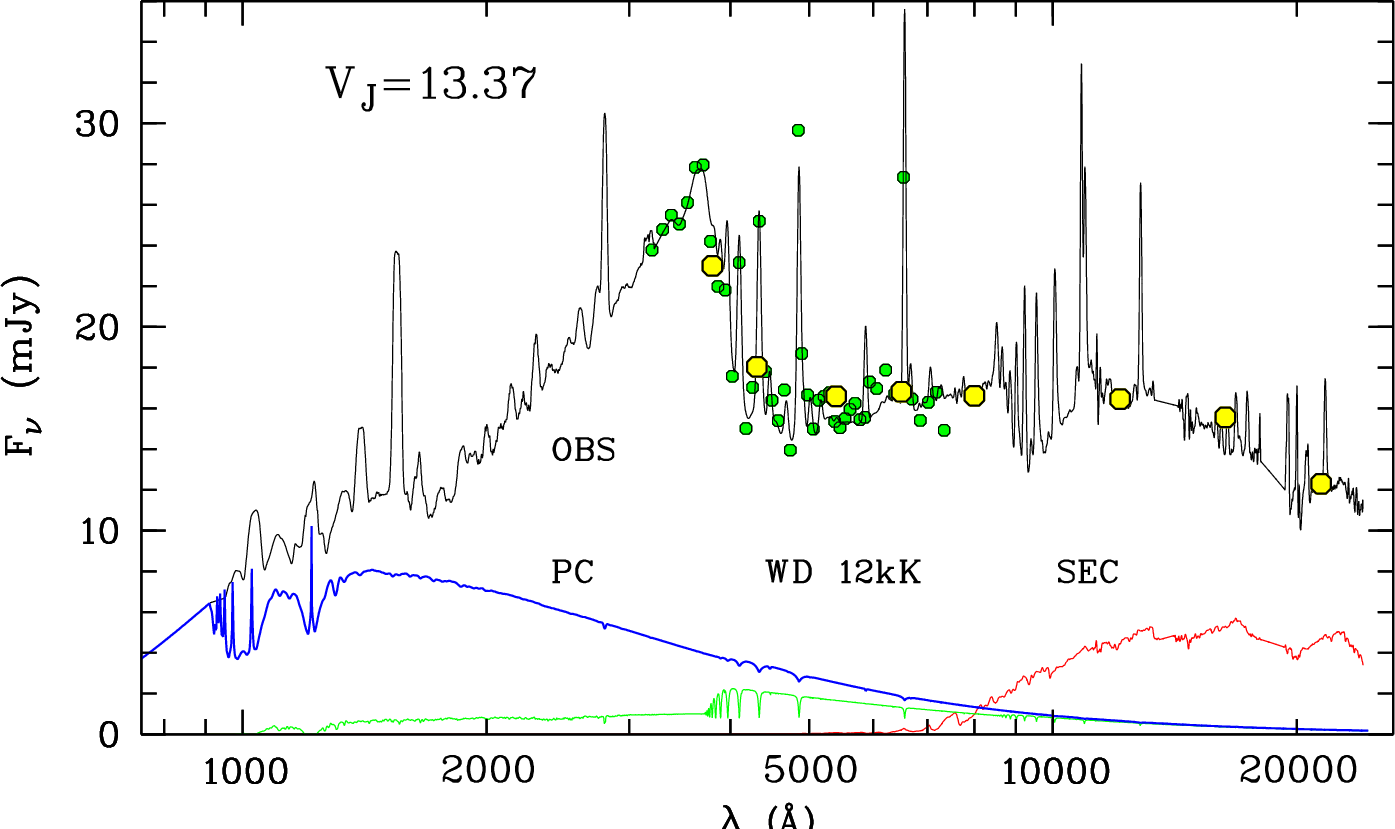}
\hfill
\includegraphics[height=74.5mm,angle=270,clip]{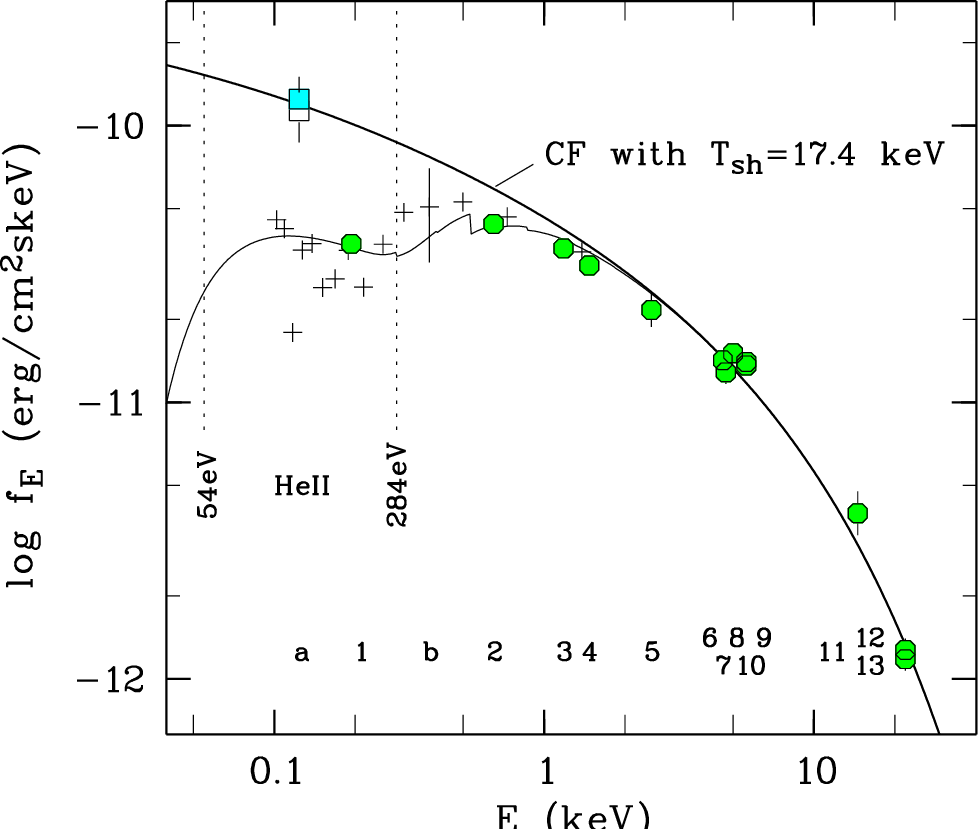}
\caption{Spin-averaged spectral energy distribution (SED) of
  \oex. Left: Ultraviolet to infrared section (smoothed, black curve),
  spectrophotometry of \citet{bathetal80} (green dots), and our
  UBVRIJHK photometry (yellow dots).  Also shown are the contributions
  of the secondary star (SEC), the intrinsic emission of the WD, and
  the radiatively heated polar caps (PC) (see text). Right: X-ray part
  of the SED. Thin black curve: Spectral model absorbed as described
  in the text. Fat black curve: Unabsorbed model. The individual
  measurements are referenced by numbers and letters: (1) ROSAT PSPC
  $0.10-0.28$\,keV (HEASARC, footnote\,2), (2) Swift $0.3-1.0$\,keV
  (HEASARC), (3) ROSAT PSPC $0.4-2$\,keV (HEASARC), (4) Swift
  $1-2$\,keV (HEASARC), (5) EINSTEIN SSS $0.5-4.5$\,keV
  \citep{singhswank93}, (6) XMM-Newton $0.3-10$\,keV
  \citep{pekoenbalman11}, (7) ASCA $0.5-10$\,keV
  \citep{ishidaetal94a}, (8) EXOSAT ME $1-10$\,keV
  \citep{rosenetal88}, (9) Swift $2-10$\,keV (HEASARC), (10) GINGA
  $2-10$\,keV \citep{ishidaetal94b}, (11) GINGA $10-20$\,keV
  \citep{ishidaetal94b}, (12) Suzaku $12-40$ keV \citep{yuasaetal10},
  and (13) NuSTAR $12-40$\,keV \citep{lunaetal18}. (a)~Cyan blue
  square: The EUV flux required to explain the observed
  \ion{He}{ii}$\lambda$4686 line flux by the Zanstra method (see
  text). (b) Crosses: The adjusted EXOSAT LE grating spectrum
  \citep{cordovaetal85} with a typical error attached to a single
  point. }
\label{fig:sed}
\end{figure*}

\subsection{Long-term mean visual brightness of EX Hydrae}
\label{sec:long}

The visual magnitude of \oex, monitored by the AAVSO since
1954,\footnote{The number of visual observations of \oex\ in the AAVSO
  archive declined drastically since 2019 in favor of CCD
  observations.}  demonstrates that it\ displays little or no
long-term variability apart from the infrequent short
outbursts. Short-term variability is better documented by the numerous
Johnson V-band light curves stored in the AAVSO archive since 2008. On
the timescale of hours, the variability is dominated by the spin
modulation with a full amplitude of about 0.4 mag. Orbit-to-orbit
variability or drifts over weeks occasionally reach 0.3 mag. The
long-term variability is still smaller. We studied it on timescales of
years and months, using the AAVSO visual and Johnson V-band
magnitudes, respectively. In Fig.~\ref{fig:vis}, we show their
distributions, excluding in both cases the infrequent short
outbursts. The mean quiescent visual magnitude for the years
$1954-\!2018\,^{1}$ is $\langle Vis\rangle\!=\!13.00$ with a standard
deviation of 0.07 mag, that of Johnson V for 2009-2023 is
$\langle V\rangle\!=\!13.43$ with 0.06\,mag. \oex\ showed no
substantial long-term variability over the last 70 years. For the
years of overlap, $2009-2018$,
$\langle V\rangle\!-\!\langle Vis\rangle\!=\!0.37$, implying a
long-term mean of $\langle V\rangle\!=\!13.37$. Our simultaneous
phase-resolved UBVRIJHK photometry of 1982 (BR08), happens to have a
spin-averaged $V\!=\!13.37$ as well. \citet{bathetal80} obtained
spectrophotometry with a wide slit between spin phases 0.066 and 0236
that extends from 3300 to 7500\,\AA. Folding their spectrophotometry
through the Johnson V-band filter gives $V\!=\!13.44$. We corrected
the spectral flux upward to match the long-term spin-average of
$V=13.37$. This is the magnitude we used as the anchor point to
calibrate the overall spin-averaged SED.

\subsection{Spin-averaged spectral energy distribution (SED)}
\label{sec:sed}

\subsubsection{Longward of the Lyman limit}
\label{sec:sedo}

We have previously constructed the spin-averaged SED of \oex\ in its
quiescent state for wavelengths between 912\,\AA\ and 24500\,\AA\
\citep[BR08,][henceforth EBRG02]{eisenbartetal02} and present an
update here.  Fig~\ref{fig:vis} shows a smoothed version of the
spin-averaged flux-calibrated SED as the black curve labeled OBS. The
infrared-optical part is based on our 1995 spectrophotometric
observations that were extensively discussed by EBRG02. It was
slightly adjusted to match (i) Johnson V=13.37, (ii) the
spectrophotometry of \citep{bathetal80}, and (iii) our UBVRIJHK
photometry as discussed in the last section. The far-ultraviolet (FUV)
part from 912-1830\,\AA\ is based on the spin-averaged spectrum
measured in 1995 with the Hopkins Ultraviolet Telescope (HUT)
\citep{greeleyetal97}. It has an absolute flux calibration reported to
be accurate to 5\% and the ORFEUS-SPAS\,II\,\footnote{Orbiting and
  Retrievable Far and Extreme Ultraviolet Spec\-tro\-graph$-$Shuttle
  Pallet Satellite II mission.} spectra of \citet{mauche99} agree
reasonably with the HUT spectroscopy. The spectral range between 1830
and 3300\,\AA\ was filled with the 1995 ultraviolet (UV) spectra from
the International Ultraviolet Explorer (IUE) discussed by EBRG02. Of
these, the spin-averaged short-wavelength (SWP) spectrum needed only a
small adjustment to match the HUT spectral flux. The mean
long-wavelength (LWR) spectrum, for which no sensible spin average
exists, smoothly closed the gap between the SWP and the Bath et al
fluxes with an adjustement of $+\!30$\%. The fact that the spectral
components from the infrared to the FUV regimes could be readily
concatenated indicates that the combined data set represents a
reliably defined SED, which we estimate is good to $\pm\,10$\%. The
integrated spin-averaged flux between 912 and 24500\,\AA\ normalized
to $V\!=\!13.37$ is
$F_\mathrm{optuv}\!=\!\ten{(4.51\!\pm\!0.45)}{-10}$\,\ergs.  In
Fig~\ref{fig:vis}, only two components are not of current accretion
origin, the secondary star and the compressionally heated WD,
represented by the spectrum of the dM star GL473AB divided by 580
(BR08) and a model spectrum of a 12000\,K, log\,$g$=8 WD with solar
composition, respectively.

\subsubsection{X-rays}
\label{sec:sedx}

The right panel of Fig.~\ref{fig:sed} shows the X-ray SED, based on
published observations or data accessed via NASA's High Energy
Astrophysics Science Archive Research Center
(HEASARC)\,\footnote{https://heasarc.gsfc.nasa.gov/cgi-bin/W3Browse/w3browse.pl}.
The archive provides either fluxes or count rates that we converted to
fluxes using the Portable Interactive Multi Mission Simulator
(PIMMS)\,\footnote{https://heasarc.gsfc.nasa.gov/cgi-bin/Tools/w3pimms/w3pimms.pl}.
We interpreted mean fluxes or count rates of longer observations as
spin-averages, although uniform phase coverage could not always be
ascertained. The data points in the figure are marked by numbers or
letters with further information given in the caption. We used
observed fluxes and avoided model-dependent inferred fluxes corrected
for sometimes large internal column densities
\citep[e.g.,][]{rosenetal91,ishidaetal94b,yuasaetal10,hayashiishida14}.
At low photon energies, we either avoided fluxes measured within the
broad absorption dip that extends from orbital phase 0.6 to 1.0
\citep{cordovaetal85,hurwitzetal97,belleetal02,hoogerwerfetal05} or we
applied an empirical wavelength-dependent correction based on the work
of \citet{hoogerwerfetal05}. Taken together, the complete sample of
non-simultaneous X-ray observations suggests that time variability
does not exceed the 10\% level.

We fit these data with a simple spectral model that mimics a
cooling flow. It consists of a series of thermal plasma spectra with
Gaunt factor and temperatures that range from the shock temperature
\ktsh\ down to 1\,keV, with the same energy emitted per keV. This
prescription disregards post-acceleration within the cooling region
and electron and ion temperatures that are not in equilibrium.
This is not a physical model, but perfectly serves to calculate the
integrated emerging X-ray flux without indulging into the possibly
complex modeling of the internal absorption of the more energetic
X-rays. Substantial reprocessing undoubtedly occurs and the absorbed
X-ray energy reappears in the ultraviolet-optical regime discussed in
the previous section.

We corrected the data for the moderate low-energy absorption by (1) an
internal neutral absorber of column density \nhint\ and a partial
covering fraction \fpc\ and (2) an interstellar neutral absorber of
column density \nhist$\,=\!\ten{3}{18}$\,\atoms, as defined by the
narrow \lyalp\ absorption line in the HST FOS spectra of \oex\
\citep[][their Sect.5.3]{eisenbartetal02}. The X-ray spectral fit
yields \ktsh$\,=\!(17\!\pm\!2)$\,keV with
\nhint$\,=\!\ten{(8.9\!\pm\!1.6)}{20}$\,\atoms\ and
\fpc$=0.63\!\pm\!0.03$. The shock temperature corresponds to a shock
height of \hsh$\,=\!0.66\,R_1$ (see the discussion in BR24). The
unabsorbed and absorbed model fluxes are shown by the fat and thin
solid curves in Fig.~\ref{fig:sed}, respectively.  The integrated
unabsorbed energy flux is
$F_\mathrm{xray}\!=\!\ten{(2.64\!\pm\!0.27)}{-10}$\,\ergs\ for energies
between the Lyman limit and 100\,keV. The integrated internally
absorbed flux is $\ten{2.30}{-10}$\,\ergs.
These numbers agree well with previous measurements of the X-ray flux
integrated over 0.1--100\,keV: $\ten{2.6}{-10}$\,\ergs\
\citep{rosenetal91}, $\ten{2.8}{-10}$\,\ergs\ \citep{yuasaetal10}, and
$\ten{2.06}{-10}$\,\ergs\ \citep{suleimanovetal19}.

\begin{figure}[t]
\includegraphics[height=90mm,angle=270,clip]{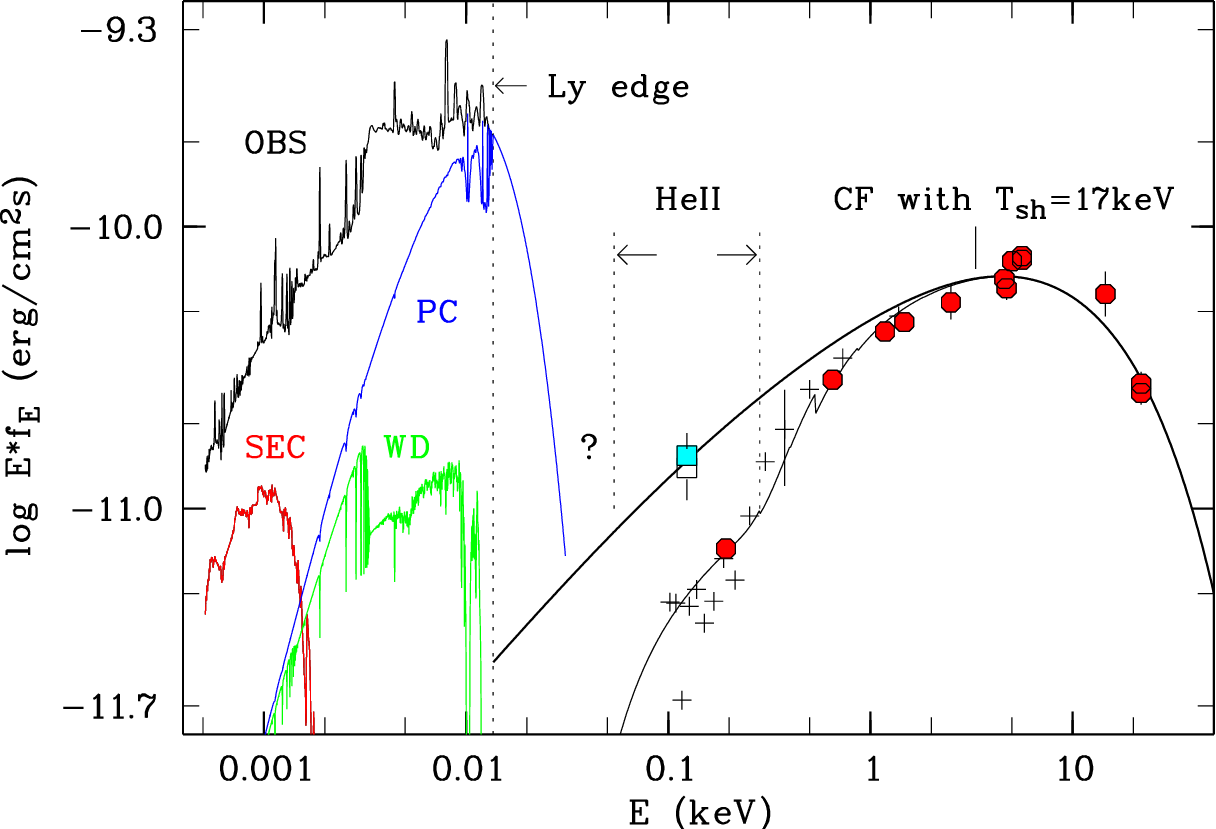}
\caption{Overall spin-averaged SED of \oex, displayed as
  log\,($E$f$_\mathrm{E}$) vs. log\,$E$. Black curve, labeled OBS:
  Observed spectrum from the K-band to the Lyman edge of
  Fig.~\ref{fig:sed}, left panel.  Red curve, labeled SEC: The
  secondary star represented by the spectrum of Gl473AB, adjusted by a
  factor of 580 (BR08). Green curve, labeled WD: Spectrum of the
  compressionally heated WD represented by a model with solar
  abundances and \teff=12000\,K (see text). Blue curve, labeled PC:
  Spectrum of the heated pole cap of the WD, represented by a WD
  spectrum of 25000\,K (see text). The X-ray part of the SED is taken
  from the right panel of Fig.~\ref{fig:sed}. }
\label{fig:sedall}
\end{figure}

\subsubsection{EUV  range}
\label{sec:sedeuv}

\begin{figure}[t]
  \includegraphics[width=89.0mm,angle=0,clip]{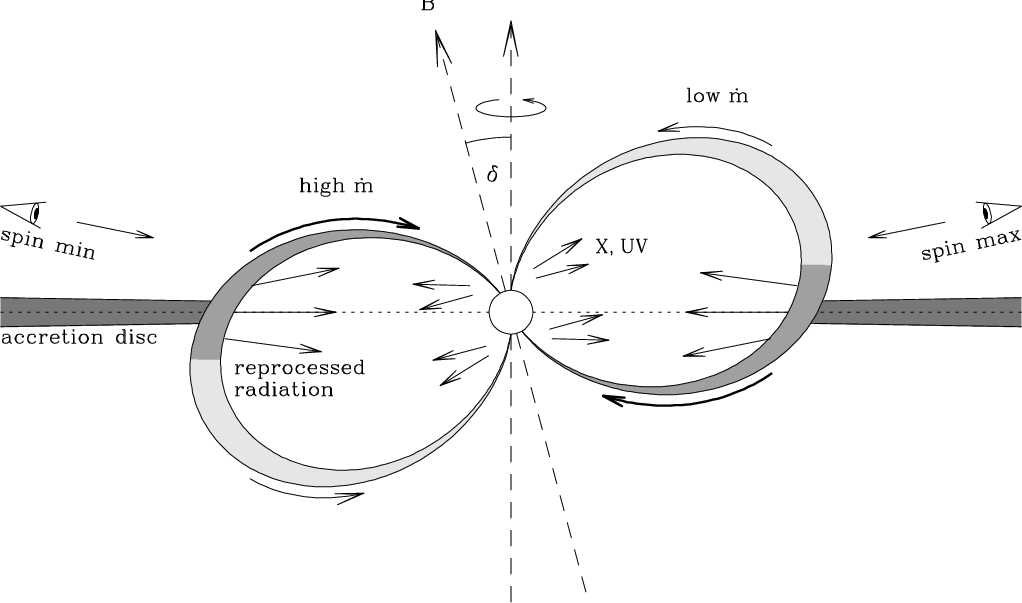}
  \caption{Schematic diagram of the magnetosphere and accretion region
    (from BR24). Viewing from the left (right) corresponds to spin
    minimum (maximum) Different shading of gray indicates different
    levels of the mass flow rate~$\dot m$. The viewing direction for
    spin maximum and spin minimum are indicated for an inclination of
    79\degr\ (from BR24). }
\label{fig:model}
\end{figure}

The transition between the FUV and the X-ray regimes in the
intervening extreme ultraviolet (EUV) region is shown in
Fig.~\ref{fig:sedall}. We discuss this region, referring to a
simplified model of the accretion geometry (Fig.~\ref{fig:model}). At
photon energies between 13.6 and $\sim\!\!90$\,eV, direct observations
of the spectral flux of \oex\ are prevented by interstellar and
internal absorption or the intrinsic faintness of the system. Of the
total unabsorbed X-ray flux of $\ten{2.64}{-10}$\,\ergs, only
$\ten{0.13}{-10}$\,\ergs\ are from the EUV regime at $E\!<\!90$\,eV.
The flux observed in the 90--155\,eV interval with the
short-wavelength (SW) spectrometer of the Extreme Ultraviolet Explorer
(EUVE) is largely lost in the noise except for a few prominent
emission lines, such as \ion{Fe}{xviii}/{xx}\,$\lambda 93.8$ and
\ion{Fe}{xx}/{xxiii}\,$\lambda 132.8$; these probably arise from the
lower accretion curtain close to the WD \citep{hurwitzetal97,belleetal02}.
This origin is consistent with the observed negative radial velocity
of the 132\,\AA\ line at spin maximum (high-$\dot m$ stream in the
upper left part of Fig.~\ref{fig:model}).

No direct measurements of the spectral flux below 90\,eV are
available. The flux between $54-284$\,eV can, however, be inferred by
the Zanstra method \citep{eisenbartetal02} that relates the
\ion{He}{ii}$\lambda4686$ photon flux to the photoionization and
subsequent recombination of \ion{He}{ii} ions by photons in the noted
energy range on the assumption that the \ion{He}{iii} region is
ionization-bounded \citep[see, e.g.,][for
details]{pattersonraymond85}. We measured the well-defined
spin-resolved \ion{He}{ii}$\lambda4686$ line fluxes in the 2004
ESO-VLT UVES\,\footnote{Ultraviolet and Visual Echelle Spectrograph at
  Unit~2 of the Very Large Telescope of the European Southern
  Observatory.} spectra of \oex\ (BR08,BR24) that were slightly
adjusted to match the presently adopted flux level. This gave a
spin-averaged mean line flux of
$f_\mathrm{4686}=\ten{1.60}{-13}$\,\ergs, taken to respresent also the
$4\pi$-average. This line flux requires a mean spectral flux in the
$54-284$\,eV interval of $\ten{1.25}{-10}$\,\ergskev, assuming
optically thin case B recombination with each \ion{He}{ii} ionizing
photon producing 0.26 \ion{He}{ii}$\lambda 4686$ photons
\citep{seaton78,pattersonraymond85}.  The result is shown as the cyan
blue square in Figs.~\ref{fig:sed} and \ref{fig:sedall} and almost
exactly equals that of the CF model. The error of $\pm\,20$\% is an
estimate. The recombination cascade produces the
\ion{He}{ii}$\lambda 1640$ ultraviolet line as well, which has a case
B line flux about 6.6 times higher than the $4686$\AA\ line
\citep{seaton78}. We measured the line flux in 36 IUE SWP spectra
(EBRG02) and found that it varied quasi-sinusoidally as a function of
spin phase, ranging from $-0.1$ to $\ten{+1.5}{-12}$\,\ergs\ between
spin minimum and spin maximum, respectively. Contrary to
\ion{He}{ii}$\lambda 4686$\AA, the UV line is noticeably affected by
internal absorption, preferably near spin minimum, when the inner
accretion region is viewed through the densest part of the stream
(Fig.~\ref{fig:model}). A quantitative interpretation requires a more
detailed study. Provisionally, we disregard the seven slightly
negative line fluxes and use the mean of the 29 positive ones that we
accept also as representative of the $4\pi$-average.  This gives a
mean EUV flux needed to produce the observed $1640$\AA\ line of
$\ten{1.13}{-10}$\,\ergskev\ (open square in Figs.~\ref{fig:sed} and
\ref{fig:sedall}, estimated error $\pm\,30$\%), consistent with, but
less reliable, than the value obtained from the $4686$\AA\ line.  The
combined evidence argues against the existence of a separate soft X-ray
component in \oex. The large internal column densities in the accretion curtain
\citep[e.g.,][]{rosenetal91,ishidaetal94b,yuasaetal10,hayashiishida14}
and the column density of $\ten{1.3}{20}$\,\atoms\ cited for the EUV
spectrum \citep{hurwitzetal97,belleetal02} suggest that the escaping
fraction is, in fact, small and the Zanstra method provides a reliable
estimate of the emission in the 54--284\,eV energy range.

Emission in the hydrogen Lyman continuum between $13.6$ and $54.4$\,eV
can neither be measured nor easily be inferred. The FUV emission
regions that are expected to contribute to the Lyman continuum are the
accretion curtain and the irradiation-heated polar cap, represented in
combination by the black curve in Figs.~\ref{fig:sed} and
\ref{fig:sedall}. EBRG02 have quantified the FUV spectral contribution
from the heated polar cap in their Fig.~9.  The corresponding HST FOS
spectral component reaches a peak flux of 7.9\,mJy at 1320\AA\ and is
best described by a 25000\,K WD model spectrum with a solar
composition and log$\,g\!=\!8$ that is modified by the
irradiation-induced fill-up of the \lyalp\ absorption line
\citep{gaensickeetal98, belleetal03}. We modeled this spectrum, using
calculations of irradiation-heated WD atmospheres by
\citet{koenigetal06}. These authors considered heating of the polar
spot in AM Her by intense cyclotron radiation, which creates a hot
corona and allows a sizable Lyman continuum to escape. In \oex, the
polar cap is instead heated by X-rays, which will affect the
atmospheric temperature stratification, but not prevent the creation
of a corona. We felt justified, therefore, to adopt a model of
\citet{koenigetal06} to represent the polar-cap emission of \oex. It
is obtained for a peak temperature of the spot of 30000\,K, an
effective radius of the emitting spot of $\ten{3.8}{8}$\,cm, and is
shown as the blue curve in Figs.~\ref{fig:sed} and \ref{fig:sedall}. A
mean temperature of 25000\,K was previously proposed by
\citet{greeleyetal97} and \citet{belleetal03} and a two-temperature
model by \citet{mauche99}.  The polar-cap emission accounts for the
major fraction of the observed emission at 912\,\AA\ and we expect a
correspondingly large contribution to the Lyman continuum at
$\lambda\!\le\!911$\AA. The curtain emission is the major component at
longer wavelengths, but drops below that of the polar-cap at
912\,\AA. The former is thought to be created in part by the
reprocessing of X-rays that hit the inner surface of the accretion
stream (Fig.~\ref{fig:model}). Given that $\tau_{911}\!>\!1$ for
\nh$\,>\!10^{17}$\,\atoms\ and the large column density presented by
the stream, escape of Lyman continuum photons from the curtain will be
restricted.  Finally, we neglected the (likely) faint helium continuum
emission between 24.6 and 54.4\,eV, marked by a '?' in
Fig. \ref{fig:sedall}.

\subsubsection{The $4\pi $-averaged energy flux of the WD}
\label{sec:4pi}

An accurate measurement of the accretion luminosity depends on the
factor, $f_{\mathrm{4\pi}}$, which converts the observed spin-averaged
flux $F_{\mathrm{obs}}$ to the $4\pi$-averaged flux
$F_{\mathrm{acc}}\!=\!f_{\mathrm{4\pi}}F_{\mathrm{obs}}$ and is often
approximated as $f_{\mathrm{4\pi}}\!\simeq\!1.0$. In \oex, this
assumption is problematic, because of the bright polar caps that are
viewed always near the horizon, given the large inclination and the
probably small tilt of the magnetic axis, $\delta\!\sim\!10$\degr\
\citep{kimbeuermann95,kimbeuermann96,allanetal98,eisenbartetal02}
(Fig.~\ref{fig:model}). To obtain an overview, we consider the simple
model of an aligned rotator emitting anisotropically as
$F\!\propto\!a_0$\,sin$\phi\,[1\!+\!2a_1\,$cos$(2\theta$)], with
azimuth (spin phase) $\phi$ and colatitude $\theta$. This leads to
$f_{\mathrm{4\pi}}\!=\!1\!+\!(2a_1)/3$, independent of $a_0$. Only
one-third of the full latitude variation $2a_1$ turns up in
$f_{\mathrm{4\pi}}$ and, for $a_1\!=\!0$, we recover the traditional
case of $f_{\mathrm{4\pi}}\!=\!1$. For the special geometry of \oex,
considering $f_{\mathrm{4\pi}}\!\ne\!1$ is likely important for the
polar caps and may be dispensable for the other emission components.
To give an example, \citet{norton93} calculated X-ray light curves of
IPs for a range of inclinations, which show a flux ratio
$F(\theta=0\degr)/F(\theta=90\degr)\!\simeq\!1.30$, implying
$a_1\!\simeq\!0.15$ and $f_{\mathrm{4\pi}}\simeq1.10$ for
$F_{\mathrm{obs}}$ given by the spin-averaged flux at
$\theta\!=\!90$\degr.

For the large polar caps in \oex, that are heated from tall accretion
curtains, we assumed a temperature varying sinusoidally between
30000\,K at the pole and 12000\,K at the equator.  The factor
$f_{\mathrm{4\pi}}$ depends on wavelength and we calculated it
numerically for the geometry of \oex\ with $i\!=\!79$\degr, assuming
blackbody emission. Table~\ref{tab:facc} lists the flux-averaged
values of $f_{\mathrm{4\pi}}$ for the respective wavelength intervals
in lines 1 to 3 of Col.~4. For the spectral components originating
from the accretion funnels and the disk, we adopted
$f_{\mathrm{4\pi}}\!=\!1.00\!\pm\!0.10$. We assumed an uncertainty of
all conversion factors of 10\% in calculating $F_{\mathrm{4\pi}}$,
adding it quadratically to the flux error.

\subsubsection{Outburst correction to $\dot M_\mathrm{1}$ }
\label{sec:ob}

The $4\pi$-averaged integrated quiescent flux in Table~\ref{tab:facc}
(line 7 of Col.~5) still needs a correction for the infrequent
outbursts that occur every about 1.5 yr and last typically 2
days. \citet{hellieretal00} found from simultaneous observations in
the 1998 outburst that $V$-band flux and the RXTE\,\footnote{Rossi
  X-ray Timing Explorer.} $3-15$\,keV X-ray flux varied similarly over
the time interval of overlap. \citet{szkodyetal02} found the same for
two outbursts of YY Dra.  Taking the V-band flux through outburst as a
template of the accretion-induced X-ray flux, \citet{hellieretal00}
estimated that the outbursts add about 4\% to the quiescent accretion
rate in \oex. We confirm this estimate as follows: The $V$-band light
curve of the 1998 outburst in Fig.~3 of \citet{hellieretal00} has an
integral flux of $\ten{3.57}{-8}$\,\ergcm\ above the quiescent level
of $V\!=\!13.37$ (or $\ten{1.62}{-14}$\,\ergs).  With one outburst
every 1.5~yr, the 3 day long 1998 outburst contributed 4.7\% of the
inter-outburst $V$-band flux integrated over 1.5 yr. Individual
outbursts last between 1 and 3 days and the more typical contribution
of a 2-day outburst would be 3.1\%. Given the remaining uncertainties
in the number of missed outbursts and the relation between the
$V$-band flux and the accretion rate, we adopted a correction to the
quiescent accretion rate of $(4\pm1)$\%.
 
\begin{table}[b]
  \caption{Integrated quiescent energy fluxes of the accretion-induced
    components (lines $1-7$) and fluxes of the secondary star and WD
    (lines 8 and 9). Col.\,3: Observed spin-averaged energy flux,
    Col.\,4: conversion factor $f_\mathrm{4\pi}$ (see text), and
    Cols.\,5 and 6: $4\pi$-averaged flux with error. All fluxes are
    in units of $10^{-11}$\,\ergs.}
\begin{tabular}{@{\hspace{1.0mm}}l@{\hspace{3.0mm}}c@{\hspace{3.0mm}}c@{\hspace{3.0mm}}c@{\hspace{3.0mm}}c@{\hspace{2.0mm}}c}
  \hline \hline \noalign{\smallskip}
  (1)       & ~~~(2)                 &   (3)  &   (4)    &  (5)   & (6)   \\
  Comp &  Band & $F_\mathrm{obs}$ & $f_\mathrm{4\pi}$ & $F_\mathrm{4\pi}$ & Error \\[0.5ex]
  \noalign{\smallskip} \hline
  \noalign{\medskip}
  Polar cap&\hspace{-1.2mm}$228-911$\,\AA&  ~~7.4  & 1.70     & 12.6   &  6.3  \\
            & $912-3650$\,\AA       &   15.7  & 1.33     & 20.9   &  2.9  \\
            & $3651-24500\,\AA$     &  ~~1.4  & 1.07     & ~~1.5  &  0.2  \\[0.5ex]
  Funnel \& disk & X-rays                &   23.0  & 1.00     & 23.0   &  3.2  \\
            & $912-3650$\,\AA       &   16.1  & 1.00     & 16.1   &  2.3  \\
            & $3651-24500\,\AA$     &  ~~8.8  & 1.00     & ~~8.8  &  1.2  \\[1.0ex]
  \multicolumn{4}{l}{\hspace{-1.2mm}Integrated accretion-induced flux $F_\mathrm{acc,q}$} & 82.9 & 9.7 \\[1.0ex]
  \noalign{\medskip} \hline
  \noalign{\medskip}
  Secondary star & bolometric            &  ~~1.0  & 1.00     & ~~1.0  &  0.1  \\
  WD 12000\,K    & bolometric            &  ~~2.1  & 1.00     & ~~2.1  &  0.1  \\[0.5ex]
  \noalign{\medskip} \hline
\end{tabular}
\label{tab:facc}
\end{table}

\subsubsection{The overall picture}
\label{sec:overall}

The integrated spin-averaged observed energy flux between 912 and
24450\,\AA\ (black curve in Fig.~\ref{fig:sed}) is
$F_{\mathrm{obs}}\!=\!\ten{45.1}{-11}$ \ergs, of which
$\ten{3.1}{-11}$\,\ergs\ are contributed by the secondary star and the
12000\,K WD (Table~\ref{tab:facc}, Col.~3, lines 8 and 9);
$\ten{17.1}{-11}$\,\ergs\ are from the spectral model of the polar
caps (lines 2 and 3). The remaining $\ten{24.9}{-11}$\,\ergs\ of the
observed flux are assigned to the accretion curtains and the disk
(lines 5 and 6). The disk itself is estimated to contribute about
$\ten{5}{-11}$\,\ergs\ or 6\% of the total flux in the bottom line of
Table~\ref{tab:facc}, which corresponds to the potential drop between
the $L_1$ point and the inner edge of the accretion disk at 9.5\,$R_1$
(BR24). Additional accretion-induced components are the integrated
observed X-ray flux between the Lyman edge and 100\,keV, corrected for
interstellar absorption (line 4), and the model-dependent estimate of
the Lyman continuum flux of the polar caps (line 1). Part of the Lyman
continuum emission is likely to be intercepted by the infalling
matter. On the other hand, the FUV emission from the funnel may have
its own Lyman continuum, not considered thus far, which may
approximately balance out. Accounting for these uncertainties, we
assign a 50\% error to this component

Of the total accretion-induced flux, less than one-third appears shortward
and 2/3 longward of 54\,eV, which marks the approximate dividing line
between original X-ray emission and largely reprocessed emissions at
longer wavelengths. This is in line with the large column densities of
internal photoelectric absorption that were employed in discussions of
high-energy X-ray observations
\citep[e.g.,][]{rosenetal88,rosenetal91,ishidaetal94b,kimbeuermann95,allanetal98,yuasaetal10,hayashiishida14}.
The principal reprocessing sites for X-rays emitted either upward or
downward are the accretion curtains and the WD atmosphere,
respectively. Minor reprocessing sites include: the bulge on the accretion
disk, the disk itself, and the secondary star. Of the reprocessed
emission, more than one half arises from X-rays emitted downward. The
efficiency of X-ray heating is supported by the lack of a hard X-ray
reflection component \citep{lunaetal18}. Furthermore, a sizeable
fraction of the upward emitted X-rays escape, as observations show.

The energy balance presented in Table~\ref{tab:facc} depends a bit on
the temperature of the compressionally heated WD. A measurement of its
temperature has not been possible so far, because accretion and X-ray
heating has never ceased in the past 70~yr of optical
coverage.\footnote{If accretion ceases, \oex\ is expected to drop to
  $B=15.8\,(15.4)$ and $V=\!15.8\,(15.5)$ for
  $T_\mathrm{wd}\!=\!12000$\,K (14000\,K), respectively.}  The
temperature expected from compressional heating is in the range of
$11000-14000$\,K \citep[][KBP11, their Tables 7 and
8]{townsleygaensicke09,palaetal22}, depending on the mean accretion
rate over the last $\sim\!10^5$\,yr.

\subsection{Current accretion rate}
\label{sec:acc}

The current accretion rate ${\dot{M}_{\mathrm{1}}}$ of \oex\ can be
obtained from the observed quiescent accretion-induced flux in Col.~5,
line 7 of Table~\ref{tab:facc} via:
\begin{equation}
  L_{\mathrm{acc}} = 4\pi d^2f_{\mathrm{ob}}F_{\mathrm{acc,q}} = (\mathrm{G}M_1/R_1\,-\mid\Phi_{L_1}\!\mid)\,\dot{M_1},
\label{eq:m1dotq}
\end{equation}
where $M_1$, $R_1$ and $d$ are from Table~\ref{tab:system},
$\Phi_{L_1}$ is the Roche potential in $L_1$, and
$f_{\mathrm{ob}}\!=\!1.04\!\pm\!0.01$ is the outburst correction. We
find $L_\mathrm{acc}\!=\!\ten{(3.33\pm0.39)}{32}$\,\erg. The
corresponding accretion rate, valid for the last seven decades, is
$\dot M_{\mathrm{1}}\!=\!\ten{(2.44\pm0.38)}{15}$\,\gs
$=\!\ten{(3.86\pm0.60)}{-11}$\,\msunyr, where the error is derived from
the observed flux, $M_1/R_1, f_{\mathrm{ob}}$, and $d$, in this
order. The derived rate agrees perfectly with the theoretical mass
transfer rate for AML by GR from Sect.~\ref{sec:trans},
$-\dot M_2\!=\!\ten{(3.90\pm0.36)}{-11}$\,\msunyr.  The agreement of
the two rates indicates that mass transfer over the last 70\,yr was
conservative and driven by GR alone. It confirms, furthermore, the
level of the sizeable spin-up contribution.  The agreement of the two
rates within their $\sim \!10\!-\!15$\% uncertainties is the best result that
has so far been obtained observationally for any CV.

Although GR is a continuous process, the actual mass transfer rate,
$\dot M_2$, may vary on a timescale of months and
years. \citet{hessmanetal00} showed that starspot activity modulates
the transfer rate through the inner Lagrangian point. With a typical
star spot cycle on the order of a decade, the effect on the mean
transfer rate over the last 70 yr is probably small.  The accretion
rate $\dot M_1$ displays less variability, because the accretion disk
acts as a buffer and the only excursions arise from the infrequent
outbursts \citep{schreiberetal00}, may they originate from thermal
disk instabilities or mass tranfer events
\citep{hellieretal89,hellieretal00,reinschbeuermann90}. More details
are given in Sect.~\ref{sec:ob}.

\subsection{Secular mean accretion rate of EX Hydrae}
\label{sec:secular}

In Sect.~\ref{sec:system}, we explain that the observed radius of the
bloated secondary star in \oex\ agrees perfectly with the prediction
of the revised (optimal) model of KBP11. This suggests that the
corresponding mass transfer rate of
$-\dot M_2\!=\!\ten{6.69}{-11}$\,\msunyr\ of the KBP11 model is
expected to represent the secular mean rate of \oex\ as well. Such a
high rate is in conflict with the current accretion rate derived in
the previous section. Since continuous or variable accretion of the
same average magnitude over a secular timescale result in the same
bloating, the obvious loophole is a process that is intermittently
active. One example is the frictional motion of the binary in slowly
expanding nova shells, as suggested by \citet{schreiberetal16}.

\section{Discussion}
\label{sec:disc}

In this paper, we obtained for the first time an accurately measured
accretion rate of a short-period CV. The derived rate of
$\dot M_1\!=\!\ten{(3.86\pm0.60)}{-11}$\,\msunyr\ agrees with the
theoretical mass transfer rate of
$-\dot M_2\!=\!\ten{(3.90\pm0.36)}{-11}$\,\msunyr\ for AML driven by
GR and enhanced by the spin-up of the WD. The latter amounts to a
sizable 31\% of the rate from GR and the measured combined rate
faithfully accounts for the one expected from theory
\citep{ritter85}. This is our principal result: over the last 70~yr,
mass transfer in \oex\ was driven by GR alone with no other AML
process contributing significantly.  We identified mass transfer as
conservative, with at most a small loss of matter from the system
\citep[e.g.,][]{kingwynn99}.

Our second major result concerns the inflated radius of the secondary
star in \oex\ that we interprete in terms of the evolutionary
calculations of KBP11. These authors showed that the evolution of
non-magnetic CVs below the period gap requires a secular AML enhanced
over GR by a factor of $f_\mathrm{GR}\!=\!(2.47\!\pm\!0.22)$, leading
to radius inflation of the secondary stars with a unique mass-radius
relation $R_2(M_2)$. KBP11 relied largely on data from
\citep{ritterkolb03}, a database that was more recently improved by
\citet{mcallisteretal19}, confirming the principle results of
KBP11. We found that the measured radius of the secondary star in
\oex, $R_2\!=\!0.1513\!\pm\!0.0022$\,\rsun, coincides with the best
fit of the revised model of KBP11 to the observed radii of
non-magnetic CVs, namely, $R_2\!=\!0.1512$\,\rsun\ at
$M_2\!=\!0.1074$\,\msun. We take this as an indication that the
secondary star in \oex\ has experienced a secular evolution closely
similar to non-magnetic CVs, with a similar AML and mass-transfer
history. If the evolution of non-magnetic CVs is, in fact,
characterized by a secular mass transfer enhanced over GR, this would
hold similarly for \oex, in constrast to its current lack of any
activity beyond GR. Hence, if the inflated radii of short-period CVs
are indeed the result of a secular mass transfer rate significantly in
excess of GR, we can exclude a continuous nature of the process (at
least for \oex\ and possibly for non-magnetic CVs as well). This is an
interesting prospect that ought to be explored with observational
proof.

We consider how definitive these conclusions are. The revised model of
KBP11 is characterized by a secular AML increased over GR by the
quoted factor $f_\mathrm{GR}\!=\!2.47\!\pm\!0.22$. The associated
higher mean mass transfer rate leads to a faster evolution (1.2
instead of 2.8 Gyr from $M_2\!=\!0.20$ to 0.10\,\msun), enhanced
compressional heating of the white dwarfs (to 13500\,K instead of
11400\,K), and increased inflation of the secondary stars, as they are
driven out of thermal equilibrium. Fitting the theoretical mass-radius
relation $R_2(M_2)$ to the observed radii of non-magnetic CVs defines
the revised model of KBP11 and yields the best-fit $f_\mathrm{GR}$
quoted above. At the same time, the standard model with AML by GR
nominally fails by 6.6 standard deviations. Both model tracks start
from a secondary star at the lower edge of the period gap with
$M_2\!=\!0.20$\,\msun\ and $R_2\!=f_\mathrm{RL}\,R_\mathrm{2,B98}$,
where $R_\mathrm{2,B98}$ is the model radius of a main-sequence star
of solar composition at an age of 5\,Gyr \citep{baraffeetal98} and
$f_\mathrm{RL}\!=\!1.045$ defines the start value of $R_2$. KBP11
derived the 4.5\% non-accretion-induced inflation by carefully
exploring the rotational and tidal deformation of Roche-lobe filling
stars and considering other effects (e.g., irradiation) that might
contribute to expansion besides mass loss. At a mass of 0.1074\,\msun,
the secondary stars in the standard and revised models have reached
radii of 0.1451\,\rsun\ and 0.1512\,\rsun, respectively, differing by
4.2\%. While the latter is obtained from the fit to the data, the
former may have a systematic error that results from uncertainties in
the stellar models, the adopted $f_\mathrm{RL}$, and intricacies of
the evolutionary calculations. An increased $f_\mathrm{RL}$, for
instance, would reduce the difference between the two evolutionary
scenarios and would not easily be dismissed, if it were the only
evidence. The concept of an increased secular AML and accretion rate,
however, has solved several long-standing problems of close-binary
evolution or led to an improved understanding. These include the
location of the period minimum, the number of period bouncers, the
orbital period distributions for CV subtypes, the nature and structure
of the period gap, the mass distribution of the WDs, their effective
tempertures, the space density of CVs, and last not least the inflated
radii of the secondary stars
\citep[KBP11,][]{bellonietal18,inightetal23b,mcallisteretal17,mcallisteretal19,palaetal17,palaetal22,schreiberetal16,schreiberetal24}.
The combined evidence strongly supports the reality of a secular AML
that is enhanced over GR. In this larger context, the interpretation
of the radius inflation of the secondary stars in short-period CVs
fits into the picture and has gained wide support \citep[see,
however,][]{littlefairetal08,sirotkinkim10}. Over the last decade, the
number of available observations has dramatically increased, in
particular, by the Sloan Digital Sky Survey
\citep[e.g.,][]{mcallisteretal19,inightetal23a}, calling for an update
of the calculations of KBP11. The nature of the added AML is still a
matter of debate and our result may contribute to a solution.

The obvious culprits for an intermittently active mass transfer are
nova outbursts, which induce a strongly enhanced AML by frictional
motion of the binary in the expanding shells every few $10^6$\,yr
\citep{schreiberetal16}. Nova outbursts are characterized by a
sufficiently short duty cycle, required by the rather homogenous
effective temperatures of the compressionally heated WDs in
non-magnetic short-period CVs; there is just a single drastic outlier,
SDSS\,J153817.35+512338.0 \citep{palaetal17,palaetal22}, which harbors
a much hotter WD than all other stars of the sample. The nova
recurrence time is short enough and the adjustment time of the stellar
radii long enough, on the order of $10^7$\,yr, that the radial
inflation proceeds smoothly in the course of the evolution. On the
other hand, the nova recurrence time exceeds the Kelvin-Helmholtz
timescale of the non-degenerate envelope of the WDs of about
$10^{5}$\,yr, suggesting that the WD temperatures may display a larger
fluctuation than the radii of the secondaries. This may, in fact, be
born out in the sample of non-magnetic CVs
\citep{mcallisteretal19,palaetal17,palaetal22}.  Hence, the nova
hypothesis does gain credibility that is quite distinct from the
independent arguments presented by \citet{schreiberetal16}.

The actual mass transfer rate in \oex\ is close to the minimal
possible one given by GR. It is tempting to think that the current
spin-up process started at synchronous rotation, about two million
years ago at the present rate. Any increased AML can only have
accelerated spin-up. It is furthermore tempting to think that the
process that disrupted synchronism was the last nova eruption. The
examples of V1500\,Cyg \citep[Nova Cygni
1975,][]{harrisoncampbell16,harrisoncampbell18} and V1674 Her
\citep[Nova Her 2021,][]{pattersonetal20,pattersonetal22} are telling
and provide insights into the possible fate of \oex.  The similarity
of the present accretion and synchronization torques in \oex\ and the
low dipolar field strength of about 0.7\,MG (BR24) indicate that
synchronism, if it ever existed, was fragile and vulnerable, quite
different from the persistence of the high-field polar V1500\,Cyg.
Currently, \oex\ is heading for spin equilibrium with
\pspin$\sim\!350$\,s, which it is expected to reach in about ten
million years.

\begin{acknowledgements}
  We thank the anonymous referee for the prompt response and helpful
  comments.  We thank, Boris G\"ansicke, Coel Hellier, Frederic
  Hessman, and Andrew Norton for discussions and comments.  Andrew
  Norton thankfully provided us with his 1993 light curve program.  We
  acknowledge with thanks the variable star observations from the
  AAVSO International Database contributed by observers worldwide and
  used in this research. We acknowledge the use of the versatile AAVSO
  on-line tool LCGv2. We acknowledge with thanks the observations
  from the Merrimack College Astronomical Research Group (MCARG)
  communicated by Christopher L. Duston.  We made use of the HEASARC
  archive and the PIMMS tool in extracting part of the X-ray data used
  in Figs.~3b and 4.
\end{acknowledgements}

\bibliographystyle{aa}

\end{document}